\documentstyle[prl,aps,preprint]{revtex}

\begin{document}
\draft

\title{
Temperature-dependent spin gap and singlet ground state in
BaCuSi$_{2}$O$_{6}$.}

\author{Y. Sasago and K. Uchinokura}

\address{Department of Applied Physics, The University of Tokyo,\\
 Bunkyo-ku, Tokyo 113, Japan.}

\author{A. Zheludev and G. Shirane}

\address{Brookhaven National  Laboratory,
Upton, NY 11973-5000, USA.}

\date{\today}

\maketitle

\begin{abstract}
Bulk magnetic measurements and inelastic neutron scattering were used to
investigate the spin-singlet ground state and magnetic gap excitations
in BaCuSi$_{2}$O$_{6}$, a quasi-2-dimensional antiferromagnet with a
bilayer structure. The results are well described by a model based on
weakly interacting antiferromagnetic dimers. A strongly
temperature-dependent dispersion in the gap modes was found. We suggest
that the observed excitations are analogous to magneto-excitons in light
rare-earth compounds, but are an intrinsic property of a simple
Heisenberg Hamiltonian for the $S=1/2$ magnetic bilayer.

\end{abstract}

\pacs{}


Quantum-disordered low-dimensional (low-D) antiferromagnets (AF) have in
the last two decades attracted a great deal of attention (see reference
lists in Refs.~\cite{Regnault94,Ma95,Xu90,Hase93B,Gopalan94}). Perhaps
the simplest spin model that has a non-magnetic ground state is of zero
dimensions. We refer here to a system composed of non-interacting dimers
of $S=1/2$ spins with AF Heisenberg intra-dimer exchange $J$, the
isolated-dimer (ID) model, where there is a finite energy gap $\Delta=J$
separating the singlet $S=0$ ground state from the excited $S=1$
triplet. One of the first compounds to which this construct was found to
be applicable was copper acetate that has been extensively studied by
G\"{u}del et al \cite{Gudel79}. More recently CaCuGe$_{2}$O$_{6}$, was
shown to possess many similar properties\cite{Sasago95}. Extended
networks of {\it interacting} dimers are currently an rapidly growing
field in quantum magnetism. This increasing interest was triggered by
recent studies of spin-ladders\cite{Gopalan94,Eccleston94,Matsuda96} and
spin-bilayers\cite{Millis94}, where inter- and intra-dimer spin-spin
interactions are usually comparable in magnitude. It is therefore very
unfortunate that the limit of weakly-coupled dimers could not be
investigated experimentally in simple model systems like copper acetate
or CaCuGe$_{2}$O$_{6}$: in the former compound inter-dimer interactions
are negligable, and only powder samples are currently available for the
latter, which prevents a detailed studie of the dispersion of the
triplet excitations.

In this work we report our studies of the spin dynamics in
BaCuSi$_{2}$O$_{6}$. We show that the system is well described by a
simple model of weakly-interacting AF dimers. Single-crystal inelastic
neutron scattering is used to measure the finite dispersion in the dimer
modes. The most intriguing and unexpected observation is that {\it the
dispersion is suppressed at temperatures higher or comparable to the
characteristic gap energy}. The physics of this behaviour could be
understood through analogy with magnetic excitons in
singlet-ground-state light rare earths.\cite{Jensen}

The tetragonal (space group $I\overline{4}m2$, $a=7.042$~\AA,
$c=11.133$~\AA) crystal structure of BaCuSi$_{2}$O$_{6}$ has been
investigated by Finger et al \cite{Finger89} and is shown schematically
in Fig.~\ref{struc}. Note that the structure is not that of previously
studied CaCuGe$_{2}$O$_{6}$ which has a zigzag-chain arrangement of Cu
sites. In BaCuSi$_{2}$O$_{6}$ the magnetic Cu$^{2+}$ ions and the
SiO$_{4}$ tetrahedra are arranged in layers parallel to the $(001)$
crystallographic plane. Within each Si-O-Cu layer the Cu$^{2+}$ ions
form a square-lattice {\it bilayer} of $S=1/2$ sites
[Fig.~\ref{struc}(a)]. The nearest-neighbor in-plane Cu-Cu distance is
large, $7.043$~\AA, equal to the $a$ lattice constant. The shortest
Cu-Cu separation, 2.72~\AA, is that between sites from complementary
planes [Fig.~\ref{struc}(b)]. As we shall see, these Cu pairs form AF
dimers. The Cu-Si-O bilayers are structurally separated from each other
by planes composed of Ba$^{2+}$ ions.

Conventional AC-SQUID measurements on single-crystal samples reveal an
activated behaviour of the magnetic susceptibility at $T\rightarrow 0$
that may be interpreted as a signature of a singlet ground state with a
spin gap. No evidence of any magnetic phase transition was found. Raw
data (not corrected for the effect of paramagnetic impurities) is
plotted in symbols in Fig.~\ref{data}(a). The observed $\chi(T)$ is in
quantitative agreement with the theoretical prediction for the ID model
with a singlet-triplet gap $\Delta=4.1(0.03)$~meV [Fig.~\ref{data}(a),
solid lines].

The dimer ground state for BaCuSi$_{2}$O$_{6}$ was confirmed in
inelastic neutron scattering experiments. These were performed on a
$10\times 4 \times 4$~mm$^{3}$ single-crystal sample grown using the
floating-zone method. The measurements were done at the High Flux Beam
reactor at Brookhaven National Laboratory on H8 and H7 3-axis
spectrometers, using a neutron beam of fixed final energy
$E_{f}=14.7$~meV with a Pyrolitic Graphite (PG) filter positioned after
the sample. PG $(002)$ reflections were used for both monochromator and
analyzer. The horizontal collimation setup was either $40'-40'-40'-80'$
or $40'-40'-80'- 80'$. The sample was mounted with the $(h,0,l)$
reciprocal-space plane coincident with the scattering plane of the
spectrometer. The single crystal was of excellent quality, with a mosaic
spread of $\approx 25'$. The measurements were done in the temperature
range $3.5-150$~K, utilizing a closed-cycle refrigerator. Inelastic
constant-$Q$ scans at $T=3.5$~K revealed the presence of a gap
excitation that appears around $(\hbar \omega)\approx 4.5$~meV
throughout the entire $(h,0,l)$ plane [Fig. \ref{data}(b)]. At all
temperatures between 3.5--150~K the inelastic peak is symmetric and has
a practically $T$-independent energy width, slightly larger than the
experimental resolution. The gap energy is in reasonably good agreement
with that deduced from susceptibility data assuming the ID model.

To identify the particulr pairs of Cu-spins that form the AF dimers we
have analyzed the $Q$-dependence of the energy-integrated intensity in
the gap excitations. Measurements at several wave vectors along
$(h,0,1)$, $(h,0,1.25)$ and $(h,0,1.5)$ show that this intensity is
independent of $h$, and only a gradual decrease is observed at large
momentum transfers. The latter may be attributed entirely to effect of
the Cu$^{2+}$ magnetic form factor $f(Q)$. In contrast, the
energy-integrated intensity is strongly dependent on the
$c^{\ast}$-component of the scattering vector, as shown in
Fig.~\ref{disp}(a). A similar periodic intensity modulation was
previously observed in copper-acetate\cite{Gudel79}. The dynamic
structure factor for isolated dimers can be obtained analytically:
\begin{equation}
S({\bf Q}, \omega)\propto \sin^{2}({\bf Qd})|f(Q)|^{2}\delta(\hbar
\omega -\Delta),
\label{sqw}
\end{equation}
where $2{\bf d}$ is the vector connecting individual spins within a
dimer. A fit of Eq.~(\ref{sqw}) to our $(0,0,l)$ data on
BaCuSi$_{2}$O$_{6}$ is shown in a solid line in Fig.~\ref{disp}(a). The
analysis immediately provides us with the vector {\bf d}: the dimers are
oriented parallel to the $c$ axis with an intradimer spin-spin
separation of 2.68(0.03)\AA. This value coincides with the
nearest-neighbor Cu-Cu distance in the crystal structure (2.73~\AA).
Note that at small $l$ the $(0.5,0,l)$ data in Fig.~\ref{disp}(a)
deviates from the theoretical curve. Indeed, preliminary experiments
indicate that a very weak inelastic peak around $\hbar \omega=4$~meV is
present in the entire $(h,k,0)$ plane, where the dimer structure factor
is zero. A finite cross-section at $l=0$ represents dynamic spin
correlations within the bilayers and thus a deviation from the
isolated-dimer model. We are now in the process of further investigating
this phenomenon. Below we concentrate only on the behaviour of the much
stronger inelastic feature at $l\ne 0$ that, as we see, is easily
understood within the framework of the dimer model.

Having established the dimerized nature of the ground state, we proceded
to study the dispersion in the dimer modes. No dispersion along the
$c^{\ast}$ direction to within experimental errosrs. This is consistent
with the layered quasi-2D structure of the material. In contrast, the
dimer excitations have a finite bandwidth along the $a^{\ast}$ direction
[Fig.~\ref{disp}(b), open circles for the 3.5~K data]. Note that the
Brillouin zone for the magnetic gap excitations is the same as for the
crystal structure, and the spin fluctuations in BaCuSi$_{2}$O$_{6}$,
unlike those in a Neel antiferromagnet, retain the periodicity of the
underlying lattice.

To analyze the observed dispersion relation we used a model Hamiltonian
for a square-lattice bilayer of $S=1/2$ spins that has been investigated
in connection to high-$T_{c}$ cuprate superconductors\cite{Millis94} and
appears to be an appropriate description for the spin arrangement in
BaCuSi$_{2}$O$_{6}$. The bilayer Hamiltonian involves two Heisenberg AF
exchange constants $J_{1}$ and $J_{2}$, between nearest neighbors from
adjacent planes that form the bilayer and between nearest neighbors
within each plane, respectively
\cite{Millis94}:

\begin{equation}
\hat{H}=J_{1}\sum_{i}\hat{\bf S}_{i}^{(1)}\hat{\bf S}_{i}^{(2)}
+ J_{2}\sum_{\langle i,j\rangle,\alpha}
\hat{\bf S}_{i}^{(\alpha)}\hat{\bf S}_{i}^{(\alpha)}
\label{Ham}
\end{equation}

Here $i$ labels sites in a given plane, $i$ and $j$ are nearest
neighbors in the same plane and $\alpha=1,2$ label the two planes
constituting the bilayer. The dispersion relation for (\ref{Ham}) can be
easily derived from Eq.~(\ref{Ham}) in the limit $J_{1}\ll J_{2}$. For
$J_{2}=0$ the dimerized ground state is exact. By treating the second
term as a perturbation the following relation can be obtained:

\begin{equation}
\hbar \omega_{{\bf Q}=(h,k,l)}=\Delta+2J_{2}\left[\cos(h)+\cos(k)\right]
\label{disp1}
\end{equation}

The form (\ref{disp1}) fits the data on BaCuSi$_{2}$O$_{6}$ very well,
as shown in solid lines in Fig.~\ref{disp}(b). For $T=3.5$~K the
least-squares refinement yields $\Delta=4.38(0.03)$~meV,
$J_{2}=0.21(0.03)$~meV and $J_{1}/J_{2}\approx 20 \ll 1$. The physical
picture for the finite-bandwidth modes are single-dimer (local)
excitations ``hopping'' from one site to another within the bilayer. In
this respect the magnetic excitations in BaCuSi$_{2}$O$_{6}$ are totally
analogous to magneto-excitons in some light rare-earth compounds
\cite{Jensen,Houmann79,Soderholm91,Zheludev96-PBANO}. The latter
excitations are single-ion crystal-field (CF) excitations ``hopping''
between rare earth centers by virtue of inter-site exchange
interactions.
 The difference is that in BaCuSi$_{2}$O$_{6}$ the localized excitations
occur within a single AF dimer, rather than on a single magnetic ion,
and therefore have an intrinsic structure factor.

The most interesting results emerge from the study of the temperature-
dependent behavior. Energy-integrated intensity was measured in
constant-$Q$ scans as a function of temperature in BaCuSi$_{2}$O$_{6}$
for ${\bf Q}=(0,0,1.5)$ and $(0.5,0,1.5)$ [Fig~\ref{vst}(a)]. The
intensity starts to decrease with increasing $T$ and goes down by
roughly a factor of four before leveling off above $T\approx 75$~K. This
behaviour is consistent with the theoretical prediction
\cite{Sasago95} for the ID-model if one uses the
previously obtained value $\Delta=4.38$~meV [Fig~\ref{vst}(a), solid
line]. What is obviously beyond the ID model and the first-order
perturbation treatment of the Hamiltonian (\ref{Ham}) is the {\it
temperature dependence of the excitation bandwidth}. At $T>50$~K the
magnitude of the dispersion along the $(h,0,1.5)$ direction is severely
reduced compared to that at 3.5~K [Fig.~\ref{disp}(a), solid circles].
The suppression of dispersion is best illustrated in [Fig.~\ref{vst}(b)]
that shows the temperature dependence of the excitation energy at ${\bf
Q}=(0,0,1.5)$ and $(0.5,0,1.5)$, where it is a minimum and maximum,
respectively. The gaps for these two wave vectors converge with
increasing $T$.

A physical understanding of the observed temperature dependence can be
drawn from the previously mentioned analogy with magneto-excitons, where
the dispersion in the magneto-exciton bands is also suppressed at high
temperatures. A simplistic picture for this is that since no two
excitations can simultaneously reside on one ion (dimer) at high
temperatures, when the excited states are thermally populated,
inter-site excitation hopping is inhibited and the bandwidth is reduced.
The RPA (Random Phase Approximation) was very successful in describing
the $T$-dependence of magnetic-excitons in Pr
metal\cite{Jensen,Houmann79,Bak75}, PrBa$_{2}$Cu$_{3}$O$_{7}$
\cite{Soderholm91}, and recently
Pr$_{2}$BaNiO$_{5}$\cite{Zheludev96-PBANO}. Within this framework at a
fixed momentum transfer the gap energy is given by\cite{Houmann79}:

\begin{equation}
\Delta(T)^{2}=\Delta_{0}^{2}-\beta\Delta_{0}R(T)
\label{RPA}
\end{equation}

Here $\Delta_{0}$ is the gap for non-interacting ions (dimers). $\beta$
is a coefficient that contains the matrix element of the orbital
momentum operator as well as the Fourier transform of the apropriate
exchange integral. In our case $\beta$ must be proportional to $J_{2}$.
$R(T)$ is a temperature-dependent renormalization factor, which is
simply the difference in thermal populations of the ground and excited
states. For singlet-to-triplet excitations we have:

\begin{equation}
R(T)=\frac{1-\exp\left(-\frac{\Delta_{0}}{T}\right)}
{1+3\exp\left(-\frac{\Delta_{0}}{T}\right)}
\label{RT}
\end{equation}

The solid lines in [Fig.~\ref{vst}(b)] are results of a fit of
Eqs.(\ref{RPA},\ref{RT}) to our data on BaCuSi$_{2}$O$_{6}$. Very good
agreement for both wave vectors is obtained with
$\Delta_{0}=4.22(0.02)$~meV and $\beta=-0.53(0.07)$~meV or
$\beta=1.15(0.07)$~meV for ${\bf Q}=(0,0,1.5)$ or ${\bf Q}=(0.5,0,1.5)$,
respectively.

As we see, by substituting CF excitations for singlet-triplet AF dimer
transitions the problem of weakly-interacting dimers may be mapped onto
the problem of weakly-interacting singlet-ground-state rare-earth ions.
In consequence the mathematical apparatus devised for magnetic excitons
in rare earths describes the behaviour of BaCuSi$_{2}$O$_{6}$ very well.
However, it is important to emphasize, that the microscopic physics of
the two systems, BaCuSi$_{2}$O$_{6}$ and light rare-earths, is totally
different. In the latter the singlet ground state and the single-ion CF
excitations are result from relativistic spin-orbit interactions. In
BaCuSi$_{2}$O$_{6}$ on the other hand, both the singlet ground state and
the temperature-dependent dispersion are an {\it intrinsic property of
the Heisenberg Hamiltonian} for the particular coupling geometry and
ratio of exchange constants.

In summary, we have observd temperature-dependend magneto-excitons in a
$S=1/2$ Heisenberg antiferromagnet.

The authors would like to thank J. M. Tranquada and P. Bak for their
interest in this work and for illuminating discussions. This study was
supported in part by NEDO (New Energy and Industrial Technology
Development Organization) International Joint Research Grant and the
U.S. -Japan Cooperative Program on Neutron Scattering. Work at
Brookhaven National Laboratory was carried out under Contract No.
DE-AC02-76CH00016, Division of Material Science, U.S. Department of
Energy.

\begin{figure} \caption{ Schematic view of the crystal structure of
BaCuSi$_{2}$O$_{6}$: the Cu-Si-O layers parallel to $(a,b)$ plane [a]
and and a projection along the $(0,1,0)$ direction [b]. Cu$^{2+}$ dimers
are arranged in a square lattice to form a bilayer of $S=1/2$ spins.}
\label{struc}
\end{figure}

\begin{figure} \caption{ (a) Magnetic susceptibility measured in a BaCuSi$_{2}$O$_{6}$
single crystal (symbols). The solid lines are theoretical curves
calculated for the isolated-dimer model. (b) Example constant-$Q$ scans
measured in BaCuSi$_{2}$O$_{6}$, showing the magnetic gap excitation at
$\hbar \omega \approx 4.5$~meV. The solid lines are Gaussian fits.}
\label{data}
\end{figure}

\begin{figure}
\caption{
(a) Measured energy-integrated intensity of the inelastic peak in
BaCuSi$_{2}$O$_{6}$ as a function of momentum transfer perpendicular to
$(a,b)$ plane. The solid line represents the structure factor for an
isolated dimer with $2d=2.68$~\AA. (b) Dispersion in the gap excitations
measured along $(h,0,1.5)$. The solid lines represent fits to the data,
as described in the text.}
\label{disp}
\end{figure}

\begin{figure}
\caption{
(a) Measured temperature dependence of the energy-integrated intensity
(a) and energy (b) of the gap excitations in BaCuSi$_{2}$O$_{6}$ at two
different wave vectors. The solid lines are fits to the data described
in the text.}
\label{vst}
\end{figure}

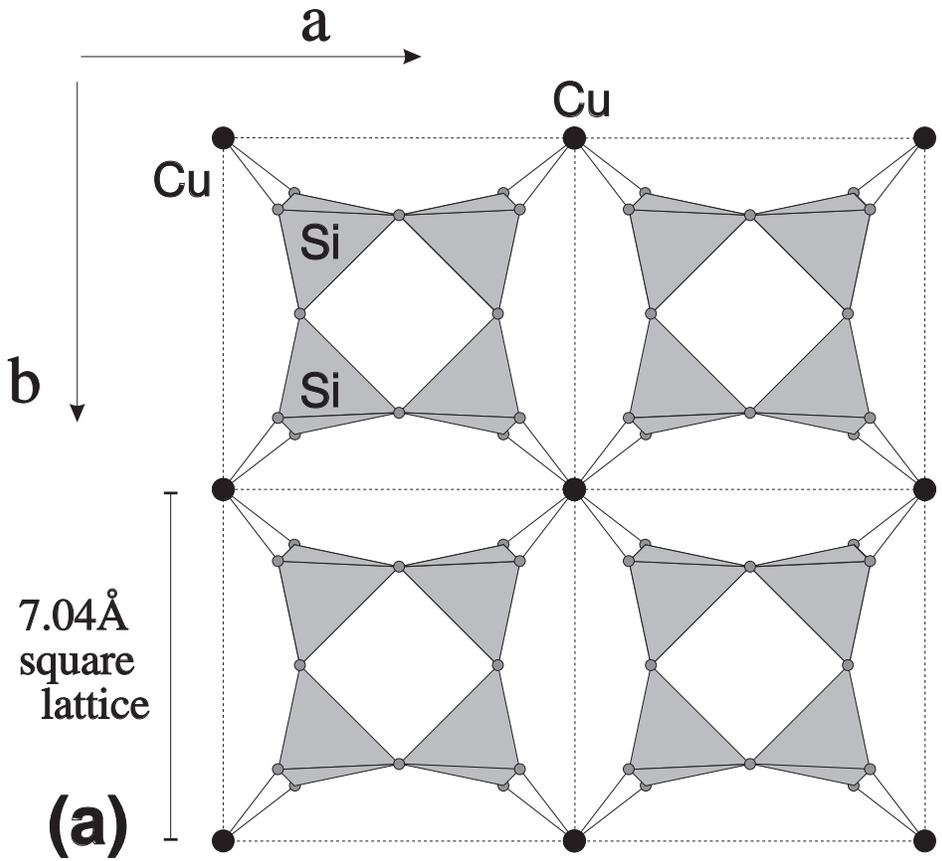
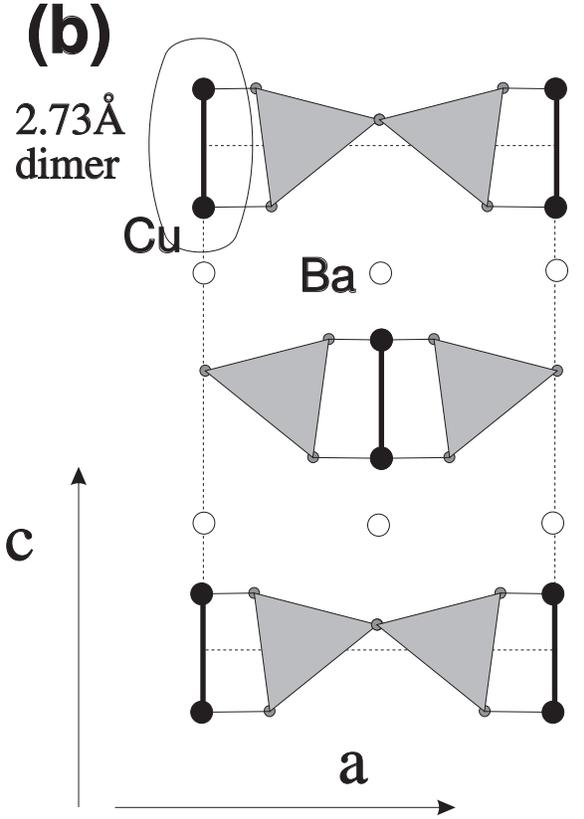
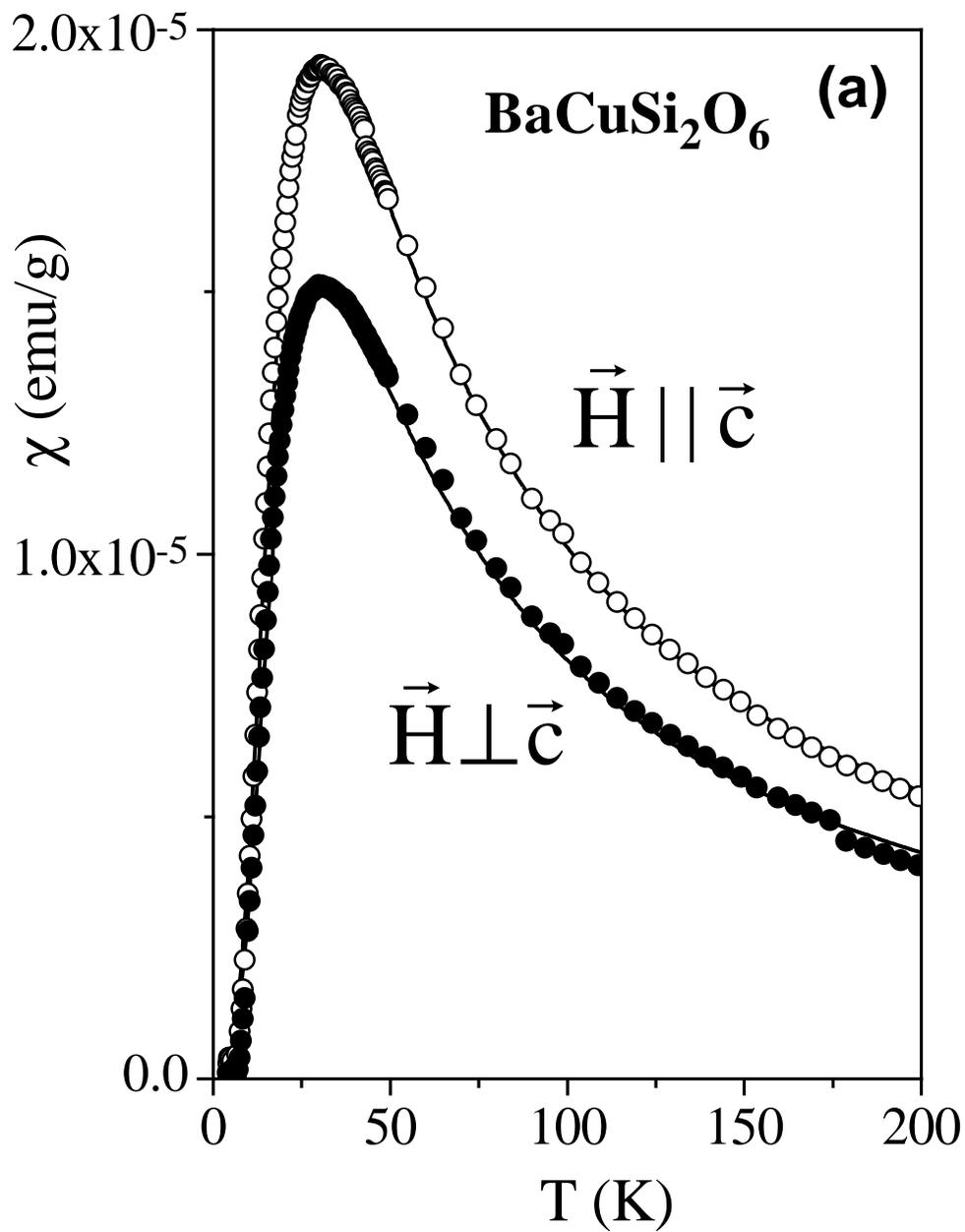
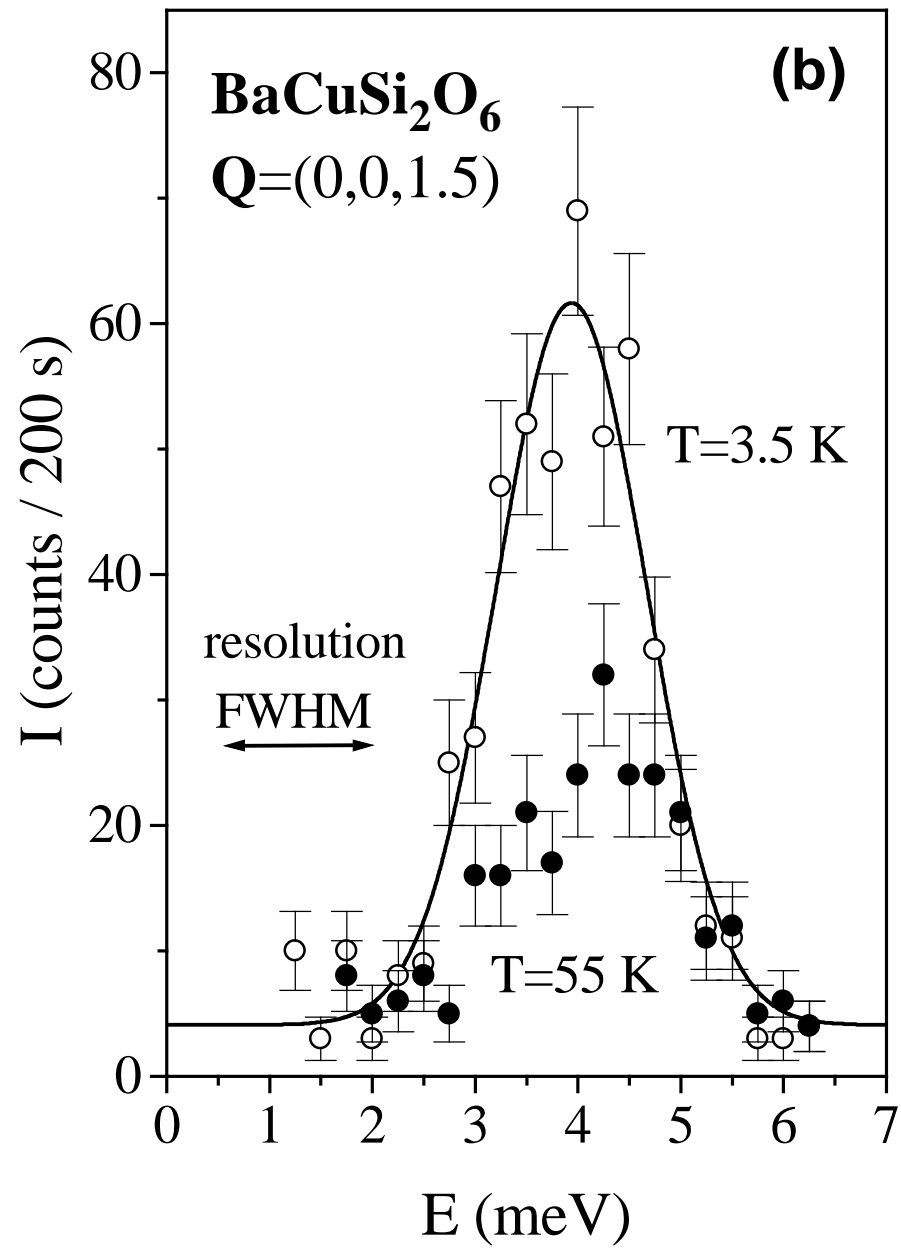
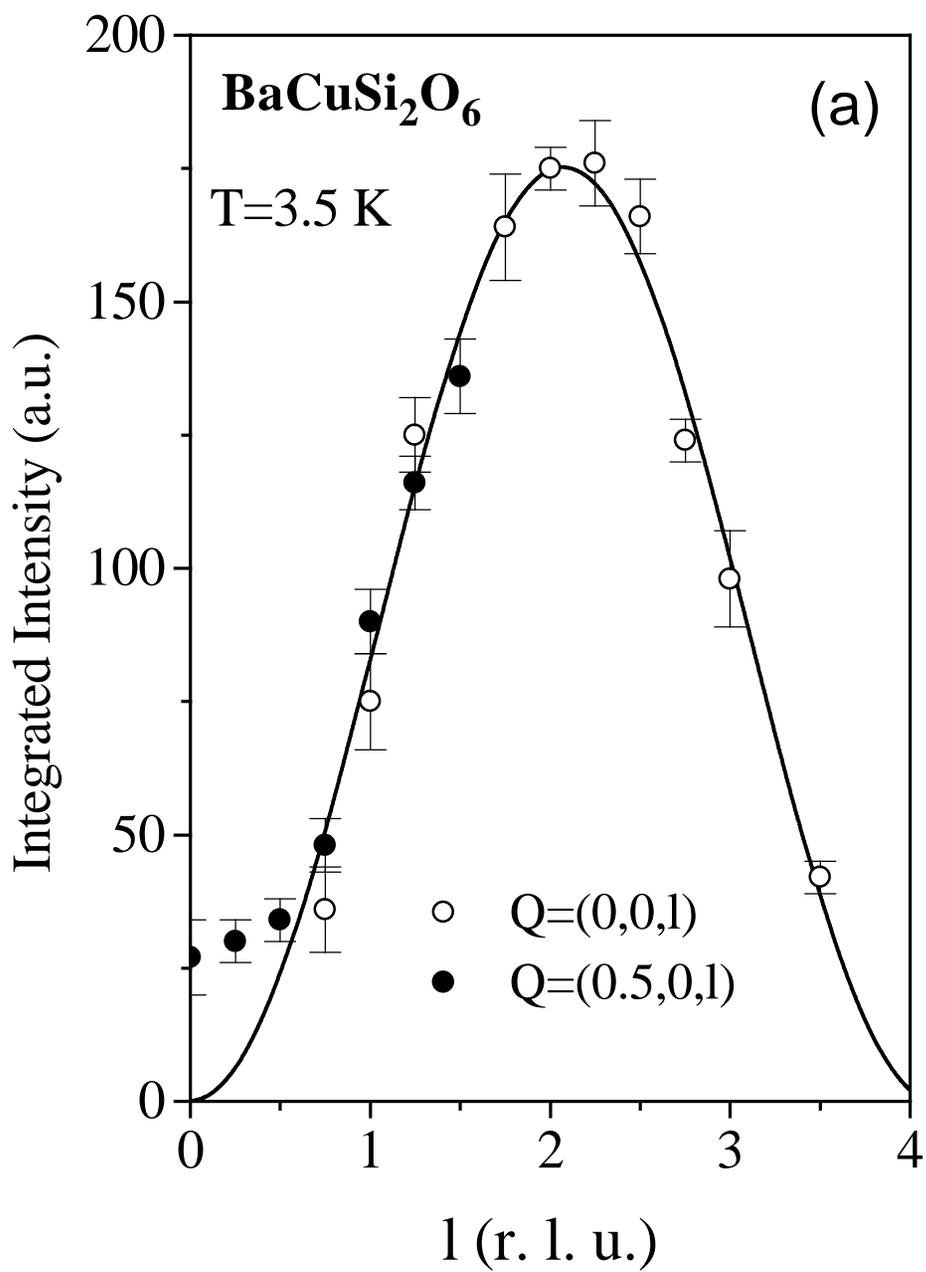
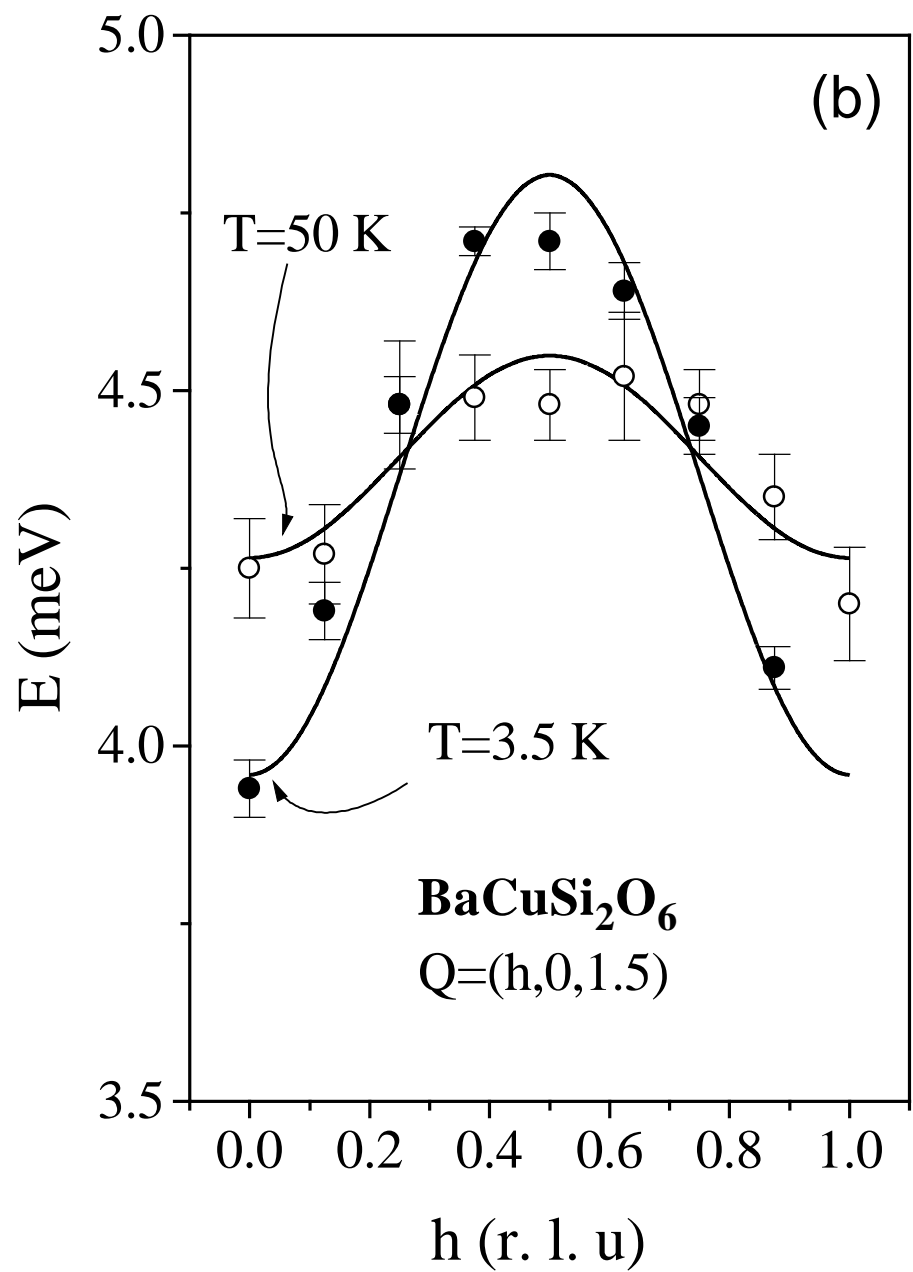
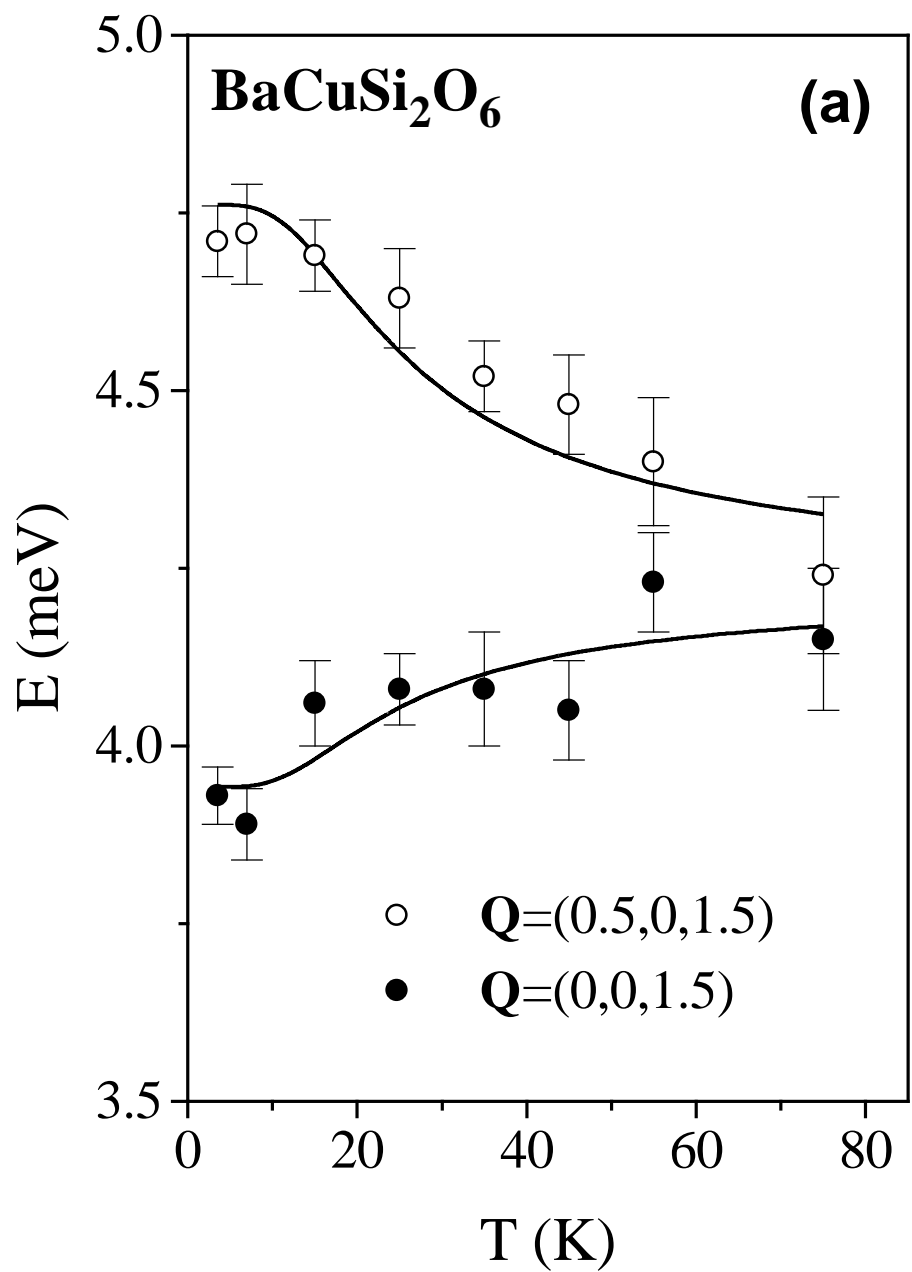
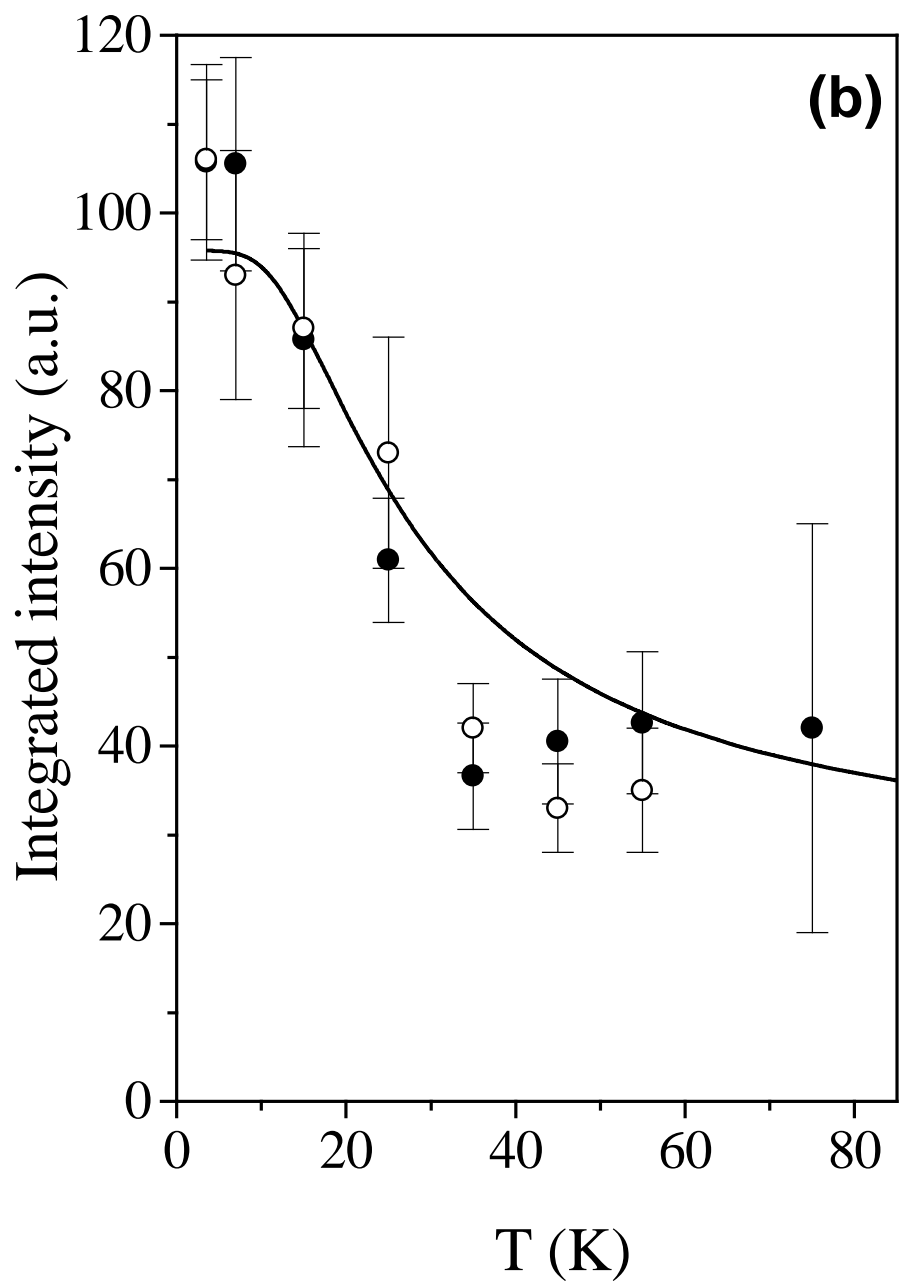
\end{document}